\documentclass[aps,showpacs,twocolumn,eqsecnum,floatfix]{revtex4}

\usepackage{epsfig}
\usepackage{latexsym}
\usepackage{mathbbol}

\begin{document}

%%%%%%%%%%%%%%%%%
%%%   TITLE   %%%
%%%%%%%%%%%%%%%%%

\title{Dynamical evolution of unstable self-gravitating scalar solitons}

\author{Miguel Alcubierre}
\email{malcubi@nuclecu.unam.mx}

\author{Jos\'e A. Gonz\'alez}
\email{cervera@nuclecu.unam.mx}

\author{Marcelo Salgado}
\email{marcelo@nuclecu.unam.mx}

\affiliation{Instituto de Ciencias Nucleares, Universidad Nacional 
Aut\'onoma de M\'exico, A.P. 70-543, M\'exico D.F. 04510 , M\'exico}

%%%%%%%%%%%%%%%%
%%%   DATE   %%%
%%%%%%%%%%%%%%%%

\date{\today}

%%%%%%%%%%%%%%%%%%%%
%%%   ABSTRACT   %%%
%%%%%%%%%%%%%%%%%%%%

\begin{abstract}
Recently, static and spherically symmetric configurations of globally
regular self-gravitating scalar solitons were found. These
configurations are unstable with respect to radial linear
perturbations.  In this paper we study the dynamical evolution of such
configurations and show that, depending on the sign of the initial
perturbation, the solitons either collapse to a Schwarzschild black
hole or else ``explode'' into an outward moving domain wall.
\end{abstract}

%%%%%%%%%%%%%%%%
%%%   PACS   %%%
%%%%%%%%%%%%%%%%


%%%%%%%%%%%%%%%%%%%%%
%%%   MAKETITLE   %%%
%%%%%%%%%%%%%%%%%%%%%

\maketitle

%%%%%%%%%%%%%%%%%%%%%%%%
%%%   INTRODUCTION   %%%
%%%%%%%%%%%%%%%%%%%%%%%%

\section{Introduction}
\label{sec:introduction}

Recently, a new family of scalar-hairy black holes (BH) and their
corresponding solitons (scalarons) were found within an Einstein-Higgs
theory with a non-positive semi-definite scalar field potential
$V(\phi)$~\cite{Nucamendi03}.  This kind of potential violates the
weak-energy condition (WEC) and therefore invalidates the
applicability of the no-scalar-hair
theorems~\cite{Heusler92,Sudarsky95,Bekenstein95}.  These configurations are
interesting in several respects. On one hand, they constitute an
example that obstructs the extension of no-hair theorems to potentials
of this type.  On the other hand, they can be useful for testing some
of the predictions of the recent isolated-horizons
formalism~\cite{Ashtekar00}. In fact, these configurations can be
shown to be unstable with respect to radial-linear perturbations, and
therefore they can be seen as {\it bound states} of non-hairy black
holes and scalarons (cf.~\cite{Ashtekar01} in the context of colored
BH). The simple perturbation analysis, however, does not provide any
definite answer about the final fate of these configurations.
Nevertheless, an heuristic analysis based on energetic arguments does
provide some clues about their fate. Presumably, the plain
Schwarzschild BH constitutes the lower energy-mass bound (the ``ground
state'') of possible BH configurations with fixed boundary conditions,
which correspond to fixed horizon area $A_h$ and asymptotic flatness.
Therefore, among all BH configurations within the theory, the
Schwarzschild BH is the energetically preferred one.

The aim of this paper is to perform a fully non-linear numerical
evolution of the scalar solitons, preparing the way for a future
analysis of the scalar-hairy black holes.  The philosophy of our
analysis is similar to the one of Straumman and Zhou for the case of
``colored solitons'' (solitons in Einstein-Yang-Mills
theory)~\cite{Zhou91}.  The initial conditions correspond to
unstable scalar solitons in a globally regular space-time. Two
different sets of initial perturbations will be considered: One that
leads to the formation of a Schwarzschild BH accompanied with a small
amount of radiated scalar field, and another one which corresponds to
an ``exploding'' configuration where a global phase transition is
triggered through the formation of an outward moving domain wall.

This paper is organized as follows.  In Section~\ref{sec:equations} we
derive the system of evolution equations and
constraints. Section~\ref{sec:initial} describes the initial
conditions corresponding to a static soliton with a small
perturbation.  We discuss the slicing conditions we use for our
simulations in Section~\ref{sec:slicing}.  Section~\ref{sec:numerics}
describes our numerical techniques.  In Section~\ref{sec:results} we
show the numerical results for the two types of 
perturbations mentioned above.  We conclude in
Section~\ref{sec:discussion}.  In the Appendix we discuss the
hyperbolicity properties of our system of evolution equation.

%%%%%%%%%%%%%%%%%%%%%%%%%%%
%%%   FIELD EQUATIONS   %%%
%%%%%%%%%%%%%%%%%%%%%%%%%%%

\section{Field equations}
\label{sec:equations}

We will consider a model of a scalar field minimally coupled to
gravity and with a non-trivial self-interaction potential. The model
is described by the Lagrangian (we will use units such that $G=c=1$):
\begin{equation}
{\cal L} = \sqrt{-g} \left[ { 1\over 16\pi } R 
- {1\over 2} \nabla_\alpha \phi \nabla^\alpha \phi 
- V(\phi) \right]  \; .
\label{eq:lag} 
\end{equation}

\begin{figure}
\vspace{5mm}
\epsfig{file=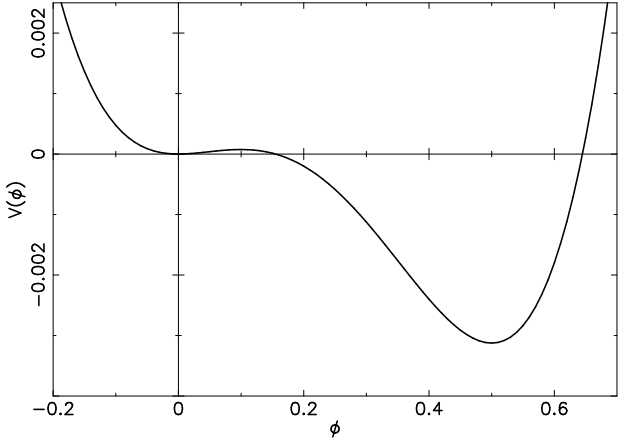,width=5.8cm,angle=270}
\caption{Scalar potential $V(\phi)$ corresponding to
Eq.~(\ref{eq:potential}) with parameters $\sigma = 1$, $\eta_1 = 0.5$,
$\eta_2 = 0.1$ and $a = 0$.}
\label{fig:potential} 
\end{figure}

We choose the following asymmetric scalar-field potential leading to
the desired asymptotically flat solutions:
\begin{eqnarray} 
V(\phi) &=& \frac{\sigma}{4} (\phi-a)^2 \left[ \rule{0mm}{5mm} (\phi-a)^2
\right.  \nonumber \\
&-& \left. \frac{4(\eta_1 + \eta_2)}{3} (\phi-a) + 2\eta_1\eta_2 
\right]  \; ,
\label{eq:potential}
\end{eqnarray}
with $\sigma$, $\eta_1$, $\eta_2$ and $a$ constant parameters. For
this class of potentials one can easily show that, if
\mbox{$\eta_1>2\eta_2>0$}, $\phi=a$ corresponds to a local minimum,
$\phi=a+\eta_1$ to the global minimum and $\phi=a+\eta_2$ to a local
maximum. The key feature of this potential for the asymptotically flat
and static solutions to exist is that the local minimum at $\phi=a$ is
also a zero of $V(\phi)$~\cite{Nucamendi03}. The factor $\sigma$ in
front of the potential fixes the scale, so one can always take
$\sigma=1$ and just re-scale everything for a different $\sigma$
afterward.  For the simulations discussed here we will take the
following values for the parameters: $\sigma = 1$, $\eta_1 = 0.5$,
$\eta_2 = 0.1$ and $a = 0$ (see Figure~\ref{fig:potential}).

The field equations following from the Lagrangian~(\ref{eq:lag}) are
the Einstein's field equations and the the Klein-Gordon (KG) equation:
\begin{equation}
G_{\mu\nu} = 8\pi T_{\mu \nu} \;,
\qquad \Box \phi = \frac{\partial V(\phi)}{\partial\phi} \; ,
\end{equation}
The stress-energy tensor for the scalar field is
\begin{equation}
T_{\mu\nu} = \nabla_\mu \phi \nabla_\nu \phi
- g_{\mu\nu}\left[{1\over 2} \nabla_{\alpha} \phi \nabla^{\alpha} 
\phi + V(\phi)\right] \; .
\end{equation}

In order to perform a numerical analysis of the problem at hand we
shall use a 3+1 approach based on the standard ADM
equations~\cite{Arnowitt62,York79}. Moreover, we shall assume that the
shift vanishes. The evolution equations for the 3-metric ($\gamma_{i
j}$) and the extrinsic curvature ($K_{i j}$) are
\begin{eqnarray}
\partial_t \gamma_{ij} &=& - 2\alpha K_{ij}  \; , 
\label{eq:gamma_ij} \\
\partial_t K_{ij}
&=& - {\cal D}_i {\cal D}_j \alpha + \alpha \left( R_{ij} +  K K_{ij} 
\right. \nonumber \\
&-& \left. 2 K_{il} K_j^l - 8 \pi M_{i j} \right)  \; , 
\label{eq:K_ij}
\end{eqnarray}
and the Hamiltonian and momentum constraints are
\begin{eqnarray}
{\cal H} &:=& R + K^2 - K_{ij} K^{ij} - 16\pi \rho = 0 \; ,
\label{eq:Ham_Con} \\
{\cal M}_i &:=& {\cal D}_l \left(
K_{\,\,\,i}^{l} - K\delta^l_{\,\,i} \right) - 8\pi J_i = 0 \; ,
\label{eq:Mom_Con}
\end{eqnarray}
with $\alpha$ the lapse function, ${\cal D}_i$ and $R_{ij}$ the
covariant derivative and Ricci tensor associated with $\gamma_{ij}$,
\mbox{$R:={\rm tr} R_{ij}$}, $K:={\rm tr} K_{ij}$, and where the
matter sources are defined in terms of the stress-energy tensor as
\begin{eqnarray}
\rho &=& n_{\mu} n_{\nu} T^{\mu \nu} \; , \\
J_i &=& - n_{\mu} T^{\mu}_{i} \; , \\
S_{ij} &=& T_{ij} \; , \\
M_{ij} &=& S_{ij} + \frac{1}{2} \; \gamma_{ij} \left( \rho
- S \right) \; ,
\label{eq:matter}
\end{eqnarray}
with $n^\mu$ the unit normal to the spatial hypersurfaces.

We shall focus on the dynamics of a spherically symmetric space-time
described by the metric
\begin{equation}
ds^2 = - \alpha^2 dt^2 + A dr^2 + B 
r^2\left(d\theta^2 + \sin^2 \theta d\varphi^2\right) \; .
\label{eq:metric}
\end{equation}
The spherical symmetry implies that all dynamical functions depend
only on $r$ and $t$.

To write down our evolution equations we start by defining the
quantities
\begin{eqnarray}
D_A &:=& \partial_r \ln A \; , \\
D_B &:=& \partial_r \ln B \; , \\
D_{\alpha} &:=& \partial_r \ln \alpha \; .
\label{eq:defD}
\end{eqnarray}
We will work with the mixed components of the extrinsic curvature $K_A
:= K_r^r \, , K_B := K^\theta_\theta=K^\phi_\phi$, and with the matter
variables $J_A := J_r$, $M_A := M^r_r $ and \mbox{$M_B :=
M^\theta_\theta = M^\phi_\phi$}.  We also introduce the extra
variables
\begin{eqnarray}
\widetilde{D} &:=& D_A - 2 D_B \; , \\
K &:=& {\rm tr} K \equiv K_A + 2 K_B \; ,
\end{eqnarray}
and use them instead of $D_A$ and $K_A$.

The evolution equations for $\{A,B,\widetilde{D},D_B,K,K_B\}$ can be
obtained directly from the ADM equations.  However, in order to have a
hyperbolic evolution system (see Appendix) we will remove the terms
proportional to $\partial_r D_B$ and $\partial_r K_B$ from the ADM
evolution equations for $K$ and $\widetilde{D}$, respectively, using
the constraints.

Spherical coordinates can be problematic at the origin. In order to
regularize the coordinate singularity we use the regularization
scheme described in Ref.~\cite{Alcubierre04a} and introduce the
auxiliary variable
\begin{equation}
\lambda := \frac{1}{r} \left( 1 - \frac{A}{B} \right) \; .
\label{eq:lambda}
\end{equation}
which will be promoted to an independent dynamical quantity and
evolved explicitly in time.

Since we have used the momentum constraint to modify the evolution
equation for $\widetilde{D}$, for regularizing the equations we need
to replace this variable with
\begin{equation}
U := \widetilde{D} - \frac{4 B \lambda}{A} \; .
\label{eq:defU}
\end{equation}

The final set of dynamical variables is then
\begin{equation}
\{ A, B, \lambda, U, D_B,  K, K_B \} \; ,
\end{equation}
and their (regularized) evolution equations are
\begin{eqnarray}
\partial_t A &=& 2 \alpha A \left( 2 K_B - K \right) \; ,
\label{eq:Adot} \\
\partial_t B &=& - 2 \alpha B K_B \; ,
\label{eq:Bdot} \\
\partial_t \lambda &=& \frac{2 \alpha A}{B} \left[ \rule{0mm}{5mm}
\partial_r K_B + 4 \pi J_A \right. \nonumber \\
&-& \left. \frac{D_B}{2} \left( K - 3 K_B \right)  \right] ,
\label{eq:lambdadot} \\
\partial_t U &=& - 2 \partial_r ( \alpha K ) + 4 \alpha D_B 
(K - 3 K_B) \nonumber \\ 
&+& 8 \alpha \left[ D_{\alpha}K_B + \frac{\lambda B}{A} (3 K_B- K)
- 4\pi J_A \right] , \hspace{9mm}
\label{eq:Udot} \\
\partial_t D_B &=& - 2 \partial_r ( \alpha K_B ) \; ,
\label{eq:DBdot} \\
\partial_t K &=& \alpha \left[ - \frac{\partial_r D_{\alpha}}{A}
- 4 K K_B + 6 K_B^2 + K^2  \right. \nonumber \\
&+& \left. \frac{D_{\alpha}}{2 A} \left( U + \frac{4 \lambda B}{A} 
\right) - \frac{D_{\alpha}^2}{A} - \frac{2 D_{\alpha}}{A r}
\right. \nonumber \\
&-& \left. 8 \pi \left( M_A + 2 M_B
- 2 \rho \right) \rule{0mm}{5mm} \right] \; ,
\label{eq:trKdot} \\
\partial_t K_B &=&  \alpha \left[ - \frac{\partial_r D_B}{2 A}
- \frac{D_{\alpha} D_B}{2 A} + \frac{D_B}{4 A}
\left(U + \frac{4 \lambda B}{A} \right)
\right. \nonumber \\
&+& \left. K K_B - 8 \pi M_B \rule{0mm}{5mm} \right] \nonumber \\
&+& \frac{\alpha}{Ar} \left[ \frac{U}{2} 
+ \frac{2 \lambda B}{A} - D_B - \lambda - D_{\alpha} \right] .
\label{eq:KBdot}
\end{eqnarray}
These equations can be easily shown to form a strongly hyperbolic
system (see Appendix).

In terms of the new variables, the Hamiltonian and momentum
constraints take the form
\begin{eqnarray}
0 &=& {\cal H} :=  \; \partial_r D_B - \frac{1}{r} \left( U - \lambda - D_B 
+ \frac{4 \lambda B}{A} \right) \nonumber \\
&-& A K_B \left( 2 K - 3 K_B \right) - D_B \left( \frac{D_B}{4} + \frac{U}{2}
+ \frac{2 \lambda B}{A} \right) \nonumber \\
&+& 8 \pi A \rho \; ,
\label{eq:ham} \\
0 &=& {\cal M} := \; \partial_r K_B - (K - 3 K_B) \left[ \frac{D_B}{2}
+ \frac{1}{r} \right] \nonumber \\
&+& 4 \pi J_A \; , \label{eq:mom}
\end{eqnarray}
and the matter variables become
\begin{eqnarray}
\rho &=& \frac{1}{2 A} \left[ \frac{\Pi^2}{B^2} + \xi^2 \right] + V \; , \\
J_A &=& - \frac{\xi \; \Pi}{A^{1/2} B} \; , \\
M_A &=& \frac{\xi^2}{A} + V \; , \\
M_B &=& V  \; .
\end{eqnarray}

Finally, the KG equation can be written as
\begin{eqnarray}
\partial_t \phi &=& \frac{\alpha \Pi}{B \sqrt{A}} \; ,
\label{eq:phidot} \\
\partial_t \xi &=& \partial_r \left( \frac{\alpha \Pi}{B \sqrt{A}} 
\right) \; ,
\label{eq:xidot} \\
\partial_t \Pi &=& - \alpha B \sqrt{A} \; \partial_{\phi} V 
+ \frac{1}{r^2} \partial_r \left( \frac{\alpha B r^2 \xi}{\sqrt{A}}
\right) \; ,
\label{eq:pidot}
\end{eqnarray}
where we have introduced the variables
\begin{equation}
\xi := \partial_r \phi \; , \qquad
\Pi := \frac{B \sqrt{A}}{\alpha} \partial_t \phi \; .
\end{equation}
For numerical purposes, the evolution equation for $\Pi$ above is
further transformed into the equivalent form:
\begin{equation}
\partial_t \Pi = - \alpha B \sqrt{A} \; \partial_{\phi} V 
+ 3 \frac{d}{dr^3} \left( \frac{\alpha B r^2 \xi}{\sqrt{A}}
\right) \; .
\label{eq:pidot2}
\end{equation}
The last term on the right hand side of this equation includes a first
derivative with respect to $r^3$.  The reason behind this
transformation is related to the regularization near the origin of the
$1/r^2$ factor in the original equation.

Our final system of evolution equations is
then (\ref{eq:Adot})-(\ref{eq:KBdot}), (\ref{eq:phidot}),
(\ref{eq:xidot}), and (\ref{eq:pidot2}).

%%%%%%%%%%%%%%%%%%%%%%%%%%%%%%
%%%   INITIAL CONDITIONS   %%%
%%%%%%%%%%%%%%%%%%%%%%%%%%%%%%

\section{Initial conditions}
\label{sec:initial}

The initial conditions used to study the evolution are the static
soliton solutions computed in Ref.~\cite{Nucamendi03}, plus/minus a
small Gaussian perturbation in $\Pi$.  Strictly speaking, pure static
initial conditions would not result in non-trivial evolution.
However, since the configurations are unstable, the truncation errors
are in fact enough to trigger the evolution.  Nonetheless, since the
numerical errors become smaller when resolution is increased, it is
better to include a finite perturbation as ``detonator''.

\subsection{Static soliton}
\label{sec:static}

As mentioned above, we shall consider small perturbations to the the
static soliton found in Ref.~\cite{Nucamendi03}. The static soliton
is obtained by first taking
\begin{eqnarray}
& B = 1 \; ,\quad D_B = K_B = K = 0 \; , \label{eq:initial1} \\
&\Pi = 0 \; . \label{eq:initial2}
\end{eqnarray}
and then solving the set of equations
\begin{eqnarray}
\partial_r A &=& A\left[\frac{1 - A}{r} + 8 \pi A r \left( \frac{\xi^2}{2 A}
+ V \right)\right] \; , \\
\partial_r \xi &=& - \xi \left(D_{\alpha} + \frac{2}{r} 
- \frac{D_A}{2} \right) + A \partial_{\phi} V \; , \\
\partial_r \alpha &=& - \frac{\alpha}{2 r} \left[ 1 - A 
- 4 \pi r^2 \left( \xi^2 - 2 A V \right)\right] \; , \\
\partial_r \phi &=& \xi \; ,
\end{eqnarray}
for $A$, $\alpha$ and $\phi$.  The first of these equations follows
from the Hamiltonian constraint~(\ref{eq:ham}), the second from the KG
equation~(\ref{eq:pidot}) and the last is just the definition of
$\xi$.  The third equation is a combination of the evolution equation
for $K_B$, Eq.~(\ref{eq:KBdot}), and the Hamiltonian constraint.  It
The equations above are solved subject to the boundary conditions
\begin{eqnarray}
&\alpha(r=0) = \alpha_0 \; , \quad A(r=0) = 1\; , \\
&\phi(r=0) = \phi_0 \; , \quad \xi(r=0) = 0  \; . &
\label{regsol}
\end{eqnarray} 

The value $\phi_0$ is used as a shooting parameter such that the
scalar-field settles asymptotically to the local minimum $V=0$ 
of the potential, thus guaranteeing asymptotic flatness.  The value
$\alpha_0$, on the other hand, is completely arbitrary.  The reason
for this is that $\alpha$ appears only in two of the equations (in one
of them through the combination $D_\alpha=\partial_r \alpha / \alpha$)
and both these equations are in fact linear in $\alpha$.  In practice
one takes $\alpha_0=1$, solves the system of equations, and later
re-scales the lapse to make sure that its asymptotic value is 1.  The
numerical code we use to solve for the initial data is fourth order
accurate, in contrast with our evolution code which is only second
order accurate (see Sect.~\ref{sec:results}).

\begin{figure}
\vspace{5mm}
\epsfig{file=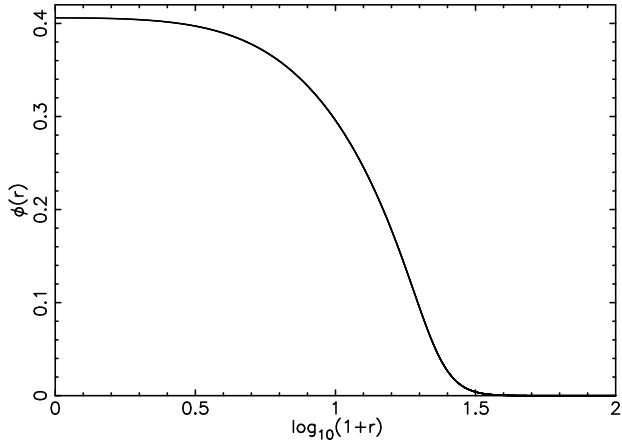,width=5.8cm,angle=270}
\caption{Scalar field $\phi$ for the static configuration (the radial
coordinate is shown on a logarithmic scale).}
\label{fig:static_phi} 
\end{figure}

\begin{figure}
\vspace{5mm}
\epsfig{file=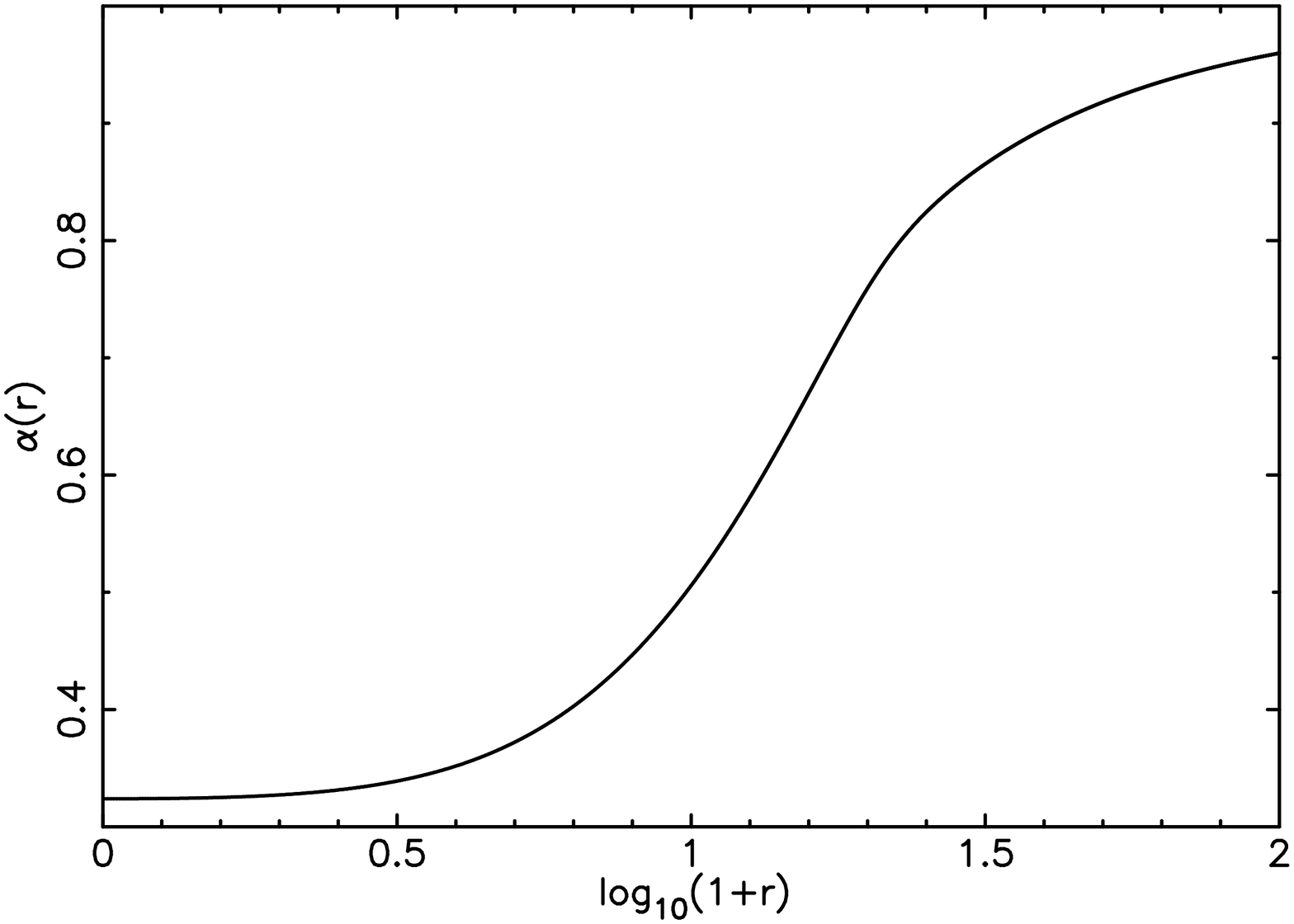,width=5.8cm,angle=270}
\caption{Same as Figure~\ref{fig:static_phi} for
the lapse function $\alpha$.}
\label{fig:static_alpha} 
\end{figure}

\begin{figure}[t]
\vspace{5mm}
\epsfig{file=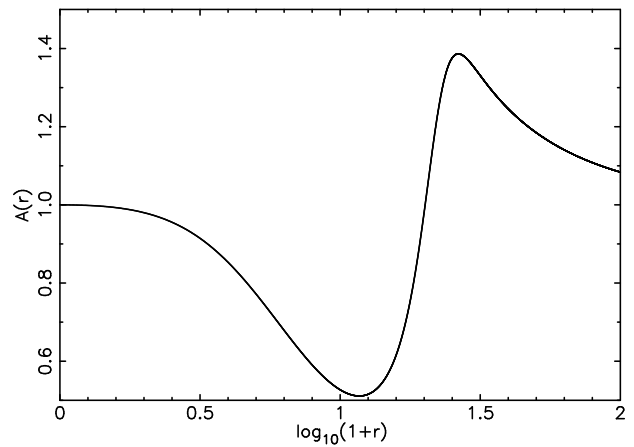,width=5.8cm,angle=270}
\caption{Same as Figure~\ref{fig:static_phi} for
the metric function $A$.}
\label{fig:static_A} 
\end{figure}

For the particular values of the parameters in the potential used
here, the static soliton solution is shown in
Figures~\ref{fig:static_phi} to \ref{fig:static_A}.  In all the plots
the numerical grid extends to $r=100$ and the radial coordinate is
shown on a logarithmic scale. The value we find for $\phi_0$ is
\begin{equation}
\phi_0 = 0.40594250807 \pm 2 \times 10^{-11} \; .
\end{equation}
The uncertainty in the eleventh significative figure is estimated by
taking the difference from the results of our two finest grids with
$\Delta r=0.025$ and $\Delta r=0.0125$. It is also important to
mention that, having fixed the parameters of the potential, this
static solution is unique, as opposed to the case of the 
corresponding hairy black holes where the family of static solutions 
is parametrized by the horizon radius \cite{Nucamendi03}.

\subsection{Perturbed soliton}
\label{sec:perturbed}

In order to perturb the static soliton we add a small Gaussian to
the time derivative of the scalar field
\begin{equation}
\Pi = \epsilon e^{-r^2/s^2} \qquad \epsilon <<1 \; ,
\label{eq:initial3}
\end{equation}
but still take $B=1$, $D_B=K=0$ (notice that it is now inconsistent to 
ask for $K_B=0$ as well). We also keep the conditions $\partial_t \Pi =
\partial_t K = 0$.  The new initial-data set of equations to be solved
is
\begin{eqnarray}
\partial_r A &=& A \left[ \frac{1 - A}{r} + 3 A r K_B^2 \right. \nonumber \\
&+& \left. 8 \pi A r \left( \frac{\Pi^2 
+ \xi^2}{2 A} + V \right) \right] \; , \\
\partial_r \xi &=& - \xi \left(D_{\alpha} + \frac{2}{r} 
- \frac{D_A}{2} \right) + A \partial_{\phi} V \; , \\
\partial^2_r \alpha &=& \partial_r \alpha \left( \frac{D_A}{2}
- \frac{2}{r} \right) \nonumber \\
&+& \alpha A \left[ 6 K_B^2 + 8 \pi \left(  \frac{\Pi^2}{A} - V \right) \right] \; , \\
\partial_r \phi &=& \xi \; , \\ 
\partial_r K_B &=& 4 \pi \, \frac{\xi \, \Pi}{A^{1/2}} - 3 \frac{K_B}{r}  \; ,
\end{eqnarray}
where the last equation arises from the fact that the momentum
constraint is no longer trivial.  Notice also that we now have a
second order equation for the lapse $\alpha$.  This is because we no
longer have the condition \mbox{$\partial_t K_B = 0$}, but rather the
condition $\partial_t K = 0$.  In the static case both conditions are
equivalent, so one can solve the simpler first order equation
corresponding to $\partial_t K_B = 0$, but in the perturbed case this
is no longer true.

The data obtained in this way will be consistent with the constraints
but will not be static, and will reduce to the static soliton solution
when $\epsilon=0$.  For the cases discussed here, we have taken the
Gaussian parameters to be $\epsilon = \pm 0.002$ and $s = 10.0$.  We
do not show plots for the initial values for $\phi$, $\alpha$ and $A$,
as they look very similar to those of the static soliton.

\subsection{Mass}
\label{sec:mass}

\begin{figure}
\vspace{5mm}
\epsfig{file=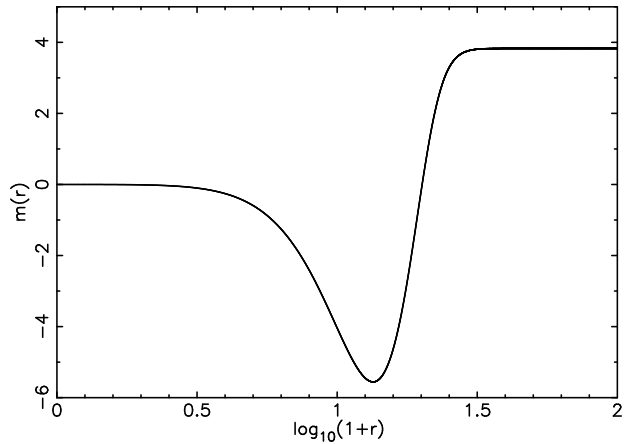,width=5.8cm,angle=270}
\caption{Same as Figure~\ref{fig:static_phi} for
the mass function $m(r)$.}
\label{fig:mass} 
\end{figure}

To find the mass of the configurations we notice that, since the
scalar field decays very rapidly with $r$, in the asymptotic region we
will have essentially the Schwarzschild metric.  If we re-parametrize
the radial metric $A$ as
\begin{equation}
A(r) = \left( 1 - \frac{2 m(r)}{r} \right)^{-1} \; , 
\end{equation}
then the mass of the configuration can be obtained from
\begin{equation}
M = \lim_{r\rightarrow\infty} m(r)
= \lim_{r\rightarrow\infty} \frac{r}{2} \left( 1 - \frac{1}{A} \right) \; .
\label{eq:mass1}
\end{equation}
In practice, there is no need to go very far, since in the region
where the scalar field is zero this expression converges very rapidly
to a limiting value.  The mass function $m(r)$ for the static
configuration is shown in Figure~\ref{fig:mass}. Its limiting value
for large $r$ turns out to be
\begin{equation}
M = 3.82719754567 \pm 2 \times 10^{-11} \; ,
\end{equation}
where again the error bar is estimated by comparing results from runs
with $\Delta r = 0.025$ and $\Delta r = 0.0125$.

For the static configuration one can show, using the Hamiltonian
constraint, that the mass given by~(\ref{eq:mass1}) can also be computed
as an integral of the energy density $\rho$ associated with the scalar
field:
\begin{equation}
M = 4 \pi \int^{\infty}_{0} \rho r^2 dr \; .
\label{eq:mass2}
\end{equation}
However, this is only true for static initial data, as extrinsic
curvature terms will enter this expression for non-static
configurations.  Using the integral expression we find the following
value for the mass of the static soliton (we have only integrated up
to the boundary of the grid, but since $\rho$ goes to zero rapidly,
the value of the integral stops changing to the last decimal figure
much before we reach the boundary)
\begin{equation}
M = 3.827197 \pm 2 \times 10^{-6} \; .
\end{equation}
The error in this value is considerably larger, which is not
surprising since the numerical integral was done using the second order
trapezium rule.  Still, it agrees with the result above to 7
significant figures.

For the perturbed configurations, the value of $M$ computed
with~(\ref{eq:mass1}) turns out to be extremely close to the value for
the static soliton, differing by about 0.1\%:
\begin{equation}
M_{\rm perturbed} = 3.8308619778 \pm 8 \times 10^{-10} \; .
\end{equation}
Notice that the mass is the same for both perturbations, which is not
surprising as $\rho$ depends only on $\Pi^2$.

%%%%%%%%%%%%%%%%%%%%%%%%%%%%%%
%%%   SLICING CONDITIONS   %%%
%%%%%%%%%%%%%%%%%%%%%%%%%%%%%%

\section{Slicing conditions}
\label{sec:slicing}

We have assumed so far a vanishing shift. We now need to specify a
slicing condition, that is, a way to find the value of the lapse
function $\alpha$ during the evolution.  A common choice is the
so-called polar slicing, which in the case of vanishing shift is
equivalent to the choice of areal (or radial) coordinates throughout
the evolution.  This consists of imposing $B(t,r)=1$, which in turn
implies $K_B(t,r)=0$, and hence $K=K_A$.  From the evolution equation
for $K_B$ one then finds an ordinary differential equation in $r$ for
the lapse that can be solved at each time step.  The main drawback of
this approach is that the radial-polar slicing gauge does not
penetrate BH horizons, since inside a horizon it is impossible to keep
the areas of spheres fixed without a non-trivial shift vector.

Since we will be interested in looking for BH horizons during the
evolution, we have decided to use horizon penetrating coordinates
instead of the above gauge.  In fact, we will use two different types
of slicing conditions, namely, harmonic slicing and maximal slicing
(see below), each adapted to the physical situation that results from
a given type of perturbation.  The adequate choice, of course, is only
known a posteriori, so in practice we have performed short trial runs
in each case with harmonic slicing in order to gain some insight about the 
dynamic behavior of the system.

Harmonic slicing is a well known condition that relates the lapse
function to the spatial volume elements: \mbox{$\alpha=f(x^i)\;
\gamma^{1/2}$}, with $\gamma$ the determinant of the spatial metric
and $f(x^i)$ an arbitrary time-independent function. This slicing 
condition is in fact 
equivalent to the requirement that the time coordinate $t$ satisfies
the wave equation $\Box t=0$, i.e. $t$ is a harmonic function.  The
harmonic condition can also be written as an evolution equation for
the lapse in the form
\begin{equation}
\partial_t \alpha = -\alpha^2 K \; ,
\label{eq:harmonic}
\end{equation}
which is a particular case of the more general Bona-Masso family of
slicing conditions~\cite{Bona94b}.

It can be shown that harmonic slicing avoids so-called ``focusing
singularities''~\cite{Bona97a,Alcubierre02b} (those for which the
spatial volume elements vanish at a bounded rate): the lapse collapses
to zero at the same rate as the volume elements.  The singularity
avoidance of harmonic slicing, however, is only marginal in the sense
that the slices come arbitrarily close to the singularity after a
finite time.  For this reason harmonic slicing is usually not used for
black hole evolutions where one wishes to avoid the singularity.

For black hole evolutions, choices different than harmonic
are usually better.  In particular, maximal slicing, defined by
$K=\partial_t K=0$, has been a standard workhorse for evolving black
holes because of its strong singularity avoidance properties
(see e.g.~\cite{Smarr78b}).  Maximal slicing leads to the following
elliptic equation for the lapse
\begin{equation}
\nabla^2 \alpha = \alpha \left[ K_{ij} K^{ij}
+ 4 \pi \left(\rho + S \right) \right] \, ,
\label{eq:maximal}
\end{equation}
In our case, we have chosen initial data that has precisely the
property that $K=\partial_t K=0$ (even in the perturbed case).
Moreover, as we will show below, one particular perturbation of the
scalar soliton leads to a BH formation, so maximal slicing is a
natural choice.

In spherical symmetry, the maximal slicing condition reduces to the
following second order ordinary differential equation for the lapse
function $\alpha$
\begin{eqnarray}
\partial_r^2 \alpha + \partial_r \alpha \left( \frac{2}{r} + D_B
- \frac{D_A}{2} \right)
= \alpha A \big[ 6 K_B^2  && \hspace{5mm} \nonumber \\
+ 4 \pi \left( \rho + S_A + 2 S_B \right) \big] .
\end{eqnarray}
This equation is solved by imposing the following boundary conditions
\begin{equation}
\partial_r \alpha \left. \rule{0mm}{4mm} \right|_{r=0} = 0 \; , \qquad
\partial_r \alpha \left. \rule{0mm}{4mm} \right|_{r=r_b} = (1-\alpha) /r \; ,
\end{equation}
with $r_b$ the position of the outer boundary.  The first condition is
just the regularity condition at the origin, and the second is a Robin
outer boundary condition that guarantees that the lapse behaves 
asymptotically as \mbox{$1+ k/r$} (with $k$ constant).

Now, although maximal slicing is well adapted for the case when the
scalar field collapses to a BH, for the ``exploding'' configuration
(i.e. an outward moving domain wall), the harmonic slicing condition
is much better suited~(\ref{eq:harmonic}).  This is because, as shown
in Sec.~\ref{sec:results} below, in that case the spacetime inside the
outward moving wall behaves in a manner similar to an anti-de-Sitter
spacetime, as the scalar field moves toward the true minimum of the
potential. This behavior will produce a big-crunch type singularity in
a finite proper time, to which maximal slicing responds by making the
lapse collapse extremely rapidly everywhere, thus completely freezing
the evolution~\cite{Hertog03a,Alcubierre04b}.  Harmonic slicing, on the 
other hand, makes the lapse collapse much slower, allowing a much longer 
evolution.  As mentioned above, such slicing allows the hypersurfaces 
to move arbitrarily close to the singularity.

%%%%%%%%%%%%%%%%%%%%%%%%%%%%%%%%%
%%%   NUMERICAL METHODOLOGY   %%%
%%%%%%%%%%%%%%%%%%%%%%%%%%%%%%%%%

\section{Numerical methodology}
\label{sec:numerics}

For the time integration in our code we use an iterative
Crank-Nicholson scheme with 3 iterations (see
e.g.~\cite{Alcubierre99d}). Derivatives are represented by second
order centered finite differences on the radial grid.  The numerical
evolution is therefore expected to be second order accurate in both
$\Delta r$ and $\Delta t$.  We also typically take $\Delta t = \Delta
r /4$ in order to be sure that we satisfy the Courant-Friedrichs-Levy
stability condition.

%%%%%%%%%%%%%%%%%%%%%%%%%%%%%%%
%%%   BOUNDARY CONDITIONS   %%%
%%%%%%%%%%%%%%%%%%%%%%%%%%%%%%%

\subsection{Boundary conditions}
\label{sec:boundary}

From the evolution equations~(\ref{eq:Adot}), (\ref{eq:Bdot})
and~(\ref{eq:phidot}) it is clear that the metric functions $A$ and
$B$ and the scalar field $\phi$ evolve only through source terms and
can be updated point-wise all the way to the boundary of the numerical
grid.  The evolution equations for other variables, however, have
spatial derivatives in the right hand side so we require a boundary
condition.  For these variables we use an outgoing wave (Sommerfeld)
boundary condition. That is, we assume that near the boundary all
dynamical variables behave as spherical waves:
\begin{equation}
h(r,t) = \frac{u(r-vt)}{r} \; .
\label{eq:boundary1}
\end{equation}
where $v$ is the wave speed.
In practice, we do not use the boundary condition~(\ref{eq:boundary1})
as above, but rather we use it in differential form
\begin{equation}
\partial_t h + v \partial_r h + v \, \frac{h}{r} = 0 \; .
\end{equation}
We also assume that the boundary is sufficiently far from the origin for the
wave speed to be essentially unit ($\alpha$ and $A$ are close to 1). So
we actually use
\begin{equation}
\partial_t h = - \partial_r h  - \frac{h}{r} \; .
\end{equation}
We use finite differences for the above equation (consistent to second
order in both space and time) and apply it to $\{\lambda, U, D_B, K,
K_B, \xi, \Pi \}$.  When using harmonic slicing, we apply the
condition also to $D_\alpha$ and evolve $\alpha$ point-wise all the
way to the boundary.

It is important to mention that the boundary conditions just described
are in fact not compatible with the constraints.  We therefore expect
a small amount of constraint violation to be introduced at the
boundaries, which will later propagate inward.  We have checked that
this has essentially no impact on the results presented here by moving
the outer boundary to different locations and comparing the results.
Constraint preserving boundary conditions are of course possible to
implement, and are certainly desirable, but they are not necessary for
the results discussed here.

%%%%%%%%%%%%%%%%%%%%%%%%%%%%
%%%   APPARENT HORIZON   %%%
%%%%%%%%%%%%%%%%%%%%%%%%%%%%

\subsection{Apparent Horizon}
\label{sec:AH}

Throughout the evolution we look for apparent horizons as an indicator
of the formation of a black hole.  Of course, an apparent horizon will
only coincide with an event horizon if the spacetime reaches a
stationary state at late times.  But if an apparent horizon is
present, an event horizon is guaranteed to exist outside of it as long
as the cosmic censor conjecture and the null energy condition
hold~\cite{Hawking73,Wald84}.  Apparent horizons have the advantage
that their definition is local in the sense that they can be located
within a given spatial hypersurface.  Event horizons, on the other
hand, are defined globally and can therefore only be located once the
whole evolution of the spacetime is known (or at least, once the
evolution is known up to the point where the spacetime is essentially
stationary).  In our case, the null energy condition is independent of
$V(\phi)$, and will therefore hold since for a null vector $\ell^\mu$
we will have $T_{\mu\nu}\ell^\mu \ell^\nu = (\ell^\mu \nabla_\mu
\phi)^2\geq 0$.  Moreover, as it turns out, when an apparent horizon
forms in our simulations the scalar field is in the region where the
potential is positive, so an event horizon is guaranteed to exist out
of it.

An apparent horizon (AH) is defined as the outermost marginally
trapped surface~\cite{Hawking73}, that is, a closed two-dimensional
surface for which the expansion of the outgoing null geodesics is
zero.  In terms of 3+1 quantities, the expansion $H$ of a congruence
of null rays moving in the outward normal direction to a given surface
takes the form~\cite{York89}
\begin{equation}
H = {\cal D}_i s^i + K_{ij} s^i s^j - K \; ,
\label{eq:expansion}
\end{equation}
with $s^i$ the unit outward pointing normal vector to the surface.  An
AH is then the outer-most closed surface such that $H=0$ everywhere on
the surface.

In spherical symmetry, this expression can be shown to reduce to the
simple form
\begin{equation}
H = \frac{1}{A^{1/2}} \left( \frac{2}{r} + D_B \right) - 2 K_B \; .
\end{equation}
In the code we evaluate this expression over the whole numerical grid
looking for places where it becomes negative (notice that for
Minkowski we simply have $H=2/r>0$).  If $H$ ever changes sign, the
outermost place where this happens is identified as the AH.

Once an apparent horizon has been located at $r=r_{AH}$, its area
will be given by:
\begin{equation}
A_{AH} = 4 \pi r_{AH}^2 B_{AH} \;
\end{equation}
where $B_{AH}$ is the metric function $B$ evaluated at
\mbox{$r=r_{AH}$}.  We can also associate a mass to this AH through
the formula
\begin{equation}
M_{AH} = \sqrt{A_{AH}/(16 \pi)} = \frac{r_{AH} \sqrt{B_{AH}}}{2} \; .
\end{equation}

%%%%%%%%%%%%%%%%%%%%%%%%%%%%%
%%%   NUMERICAL RESULTS   %%%
%%%%%%%%%%%%%%%%%%%%%%%%%%%%%

\section{Results from numerical simulations} 
\label{sec:results}

As already mentioned, the perturbations corresponding to
$\epsilon=0.002$ and $\epsilon=-0.002$ result in very different
evolutions, so we will consider each case in turn.

%%%%%%%%%%%%%%%%%%%%
%%%   COLLAPSE   %%%
%%%%%%%%%%%%%%%%%%%%

\subsection{Collapse to a black hole}
\label{sec:collapse}

We will first consider the perturbed configuration corresponding to
\mbox{$\epsilon=-0.002$}.  In this case we will use maximal slicing to
determine the lapse.  We have done numerical simulations using four
different grid resolutions, $\Delta r= (0.1,0.05,0.025,0.0125)$.  In
all cases the numerical grid extended to $r=100$.

\begin{figure}		
\vspace{5mm}
\epsfig{file=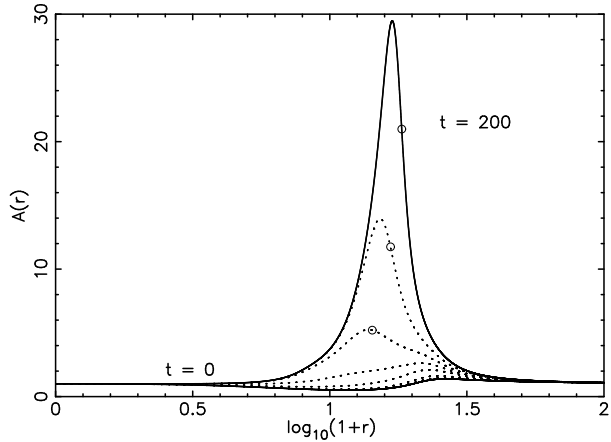,width=5.8cm,angle=270}
\caption{Evolution of the metric function $A$ for the perturbation
with \mbox{$\epsilon=-0.002$}.  Notice how at late times the metric
function grows indicating slice stretching. The circles show the
location of the apparent horizon.}
\label{fig:collapse_A} 
\end{figure}

\begin{figure}
\vspace{5mm}
\epsfig{file=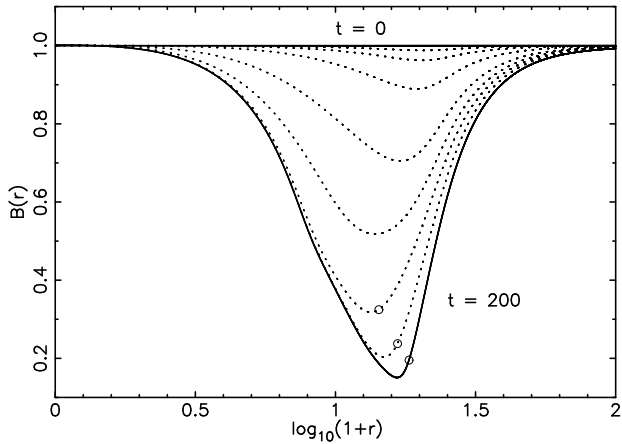,width=5.8cm,angle=270}
\caption{Evolution of the metric function $B$ for the perturbation
with \mbox{$\epsilon=-0.002$}. The circles show the
location of the apparent horizon. }
\label{fig:collapse_B} 
\end{figure}

\begin{figure}
\vspace{5mm}
\epsfig{file=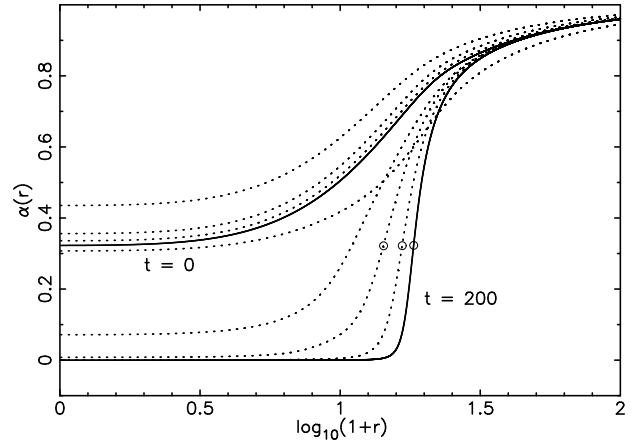,width=5.8cm,angle=270}
\caption{Evolution of the lapse function $\alpha$ for the perturbation
with \mbox{$\epsilon=-0.002$}. Notice the collapse of the lapse at
late times, indicative of the approach to a singularity. 
The circles show the location of the apparent horizon.}
\label{fig:collapse_alpha} 
\end{figure}

Figures~\ref{fig:collapse_A} and~\ref{fig:collapse_B} show the
evolution of the metric functions $A$ and $B$.  In all the plots solid
lines correspond to initial and final configurations, and dotted lines
to intermediate stages (the separation in time between these lines is
$\Delta t=25$).  At first, there are small oscillations around the
initial value, but later on the radial metric starts to grow in a form
characteristic of the slice stretching associated with BH spacetimes.
Figure~\ref{fig:collapse_alpha} shows the behavior of the lapse
function.  Again, we see the characteristic collapse of the lapse
indicative of the approach to a singularity.

\begin{figure}
\vspace{5mm}
\epsfig{file=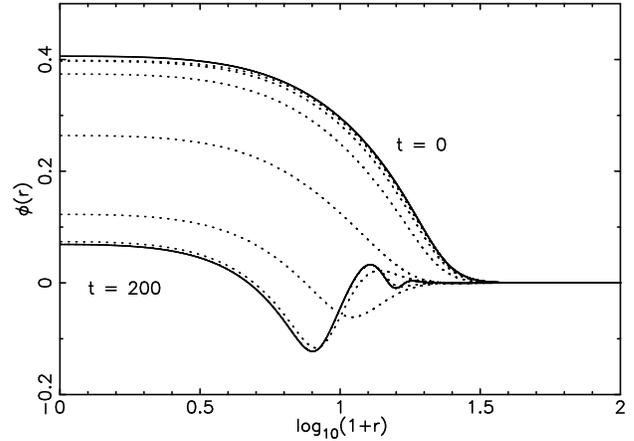,width=5.8cm,angle=270}
\caption{Evolution of the scalar field $\phi$ for the perturbation
with \mbox{$\epsilon=-0.002$}. At late times the scalar field has
values below $0.1$ everywhere, which implies that we are in the region
where the potential is positive.}
\label{fig:collapse_phi} 
\end{figure}

The corresponding evolution of the scalar field can be seen in
Figure~\ref{fig:collapse_phi}.  Notice how the scalar field moves
toward the local minimum at $\phi=0$ everywhere.  At late times, the
evolution of the inner regions is frozen due to the collapse of the
lapse there.  By that time, however, the scalar field has values below
$0.1$ everywhere, which implies that we are in the region where the
potential is positive.

\begin{figure}
\vspace{5mm}
\epsfig{file=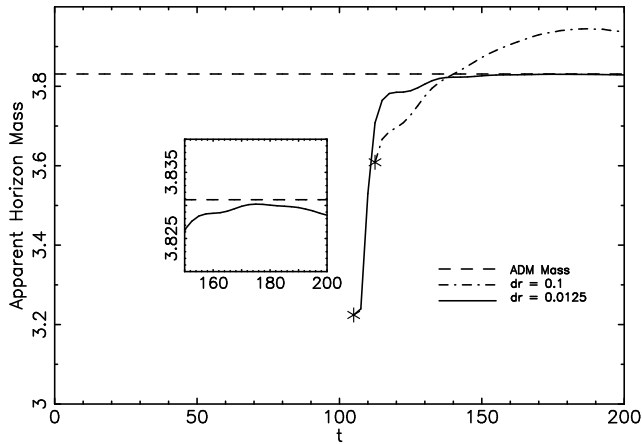,width=5.8cm,angle=270}
\caption{Apparent horizon mass for the perturbation with
\mbox{$\epsilon=-0.002$}.  An AH first appears at $t \sim 115$. The
dashed line indicates the initial ADM mass of the spacetime, and the
solid and dash-dotted lines the horizon mass for the highest and
lowest resolutions respectively. The asterisks mark the first
appearance of the apparent horizon.}
\label{fig:collapse_ah} 
\end{figure}

The slice stretching and collapse of the lapse are indicative of the
presence of a BH, but they are not enough to prove that one is
present, so we have looked for the appearance of an AH.
Figure~\ref{fig:collapse_ah} shows the mass associated with the AH's
found during the simulation.  From the figure one can see that an AH
first appears at $t \sim 115$.  The mass associated with this horizon
grows slightly at first as scalar field falls into the BH, but later
settles.  For consistency, this mass should always remain below the
initial ADM mass of the spacetime.  From the figure we see that for
low resolution the numerical error causes the horizon mass to
eventually become larger than the ADM mass.  At higher resolutions,
however, the horizon mass remains below the ADM mass throughout the
entire evolution (see inset of Fig.~\ref{fig:collapse_ah}).  
The small difference between the horizon mass and the ADM mass indicates 
that a small amount of scalar field has been radiated away.

\begin{figure}
\vspace{5mm}
\epsfig{file=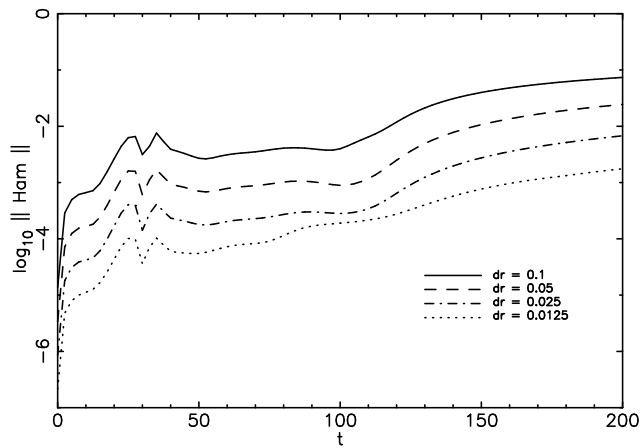,width=5.8cm,angle=270}
\caption{Logarithm of the root mean square of the Hamiltonian
constraint as a function of time for the perturbation with
\mbox{$\epsilon=-0.002$}, for four different resolutions.}
\label{fig:collapse_conv} 
\end{figure}

A crucial test of the validity of our code is the behavior of the
constraints.  Analytically, the constraints must be identically zero,
but numerical errors mean that for our simulations the constraints
have in fact finite values.  Still, those values should converge to
zero as resolution is increased.  Figure~\ref{fig:collapse_conv} shows
the logarithm of the root mean square of the Hamiltonian constraint
over the numerical grid as a function of time for the four resolutions
used in our simulations $\Delta r=(0.1, 0.05, 0.025, 0.0125)$.  As
expected, the Hamiltonian constraint is converging to zero to second
order in $\Delta r$.

%%%%%%%%%%%%%%%%%%%%%
%%%   EXPLOSION   %%%
%%%%%%%%%%%%%%%%%%%%%

\subsection{Explosion}
\label{sec:explosion}

Next we consider the perturbed configuration corresponding to
\mbox{$\epsilon=+0.002$}.  In this case we will use harmonic slicing
to determine the lapse.  We will use the same grid resolutions as
before, but will extend the numerical grid further out to $r=500$.

\begin{figure}
\vspace{5mm}
\epsfig{file=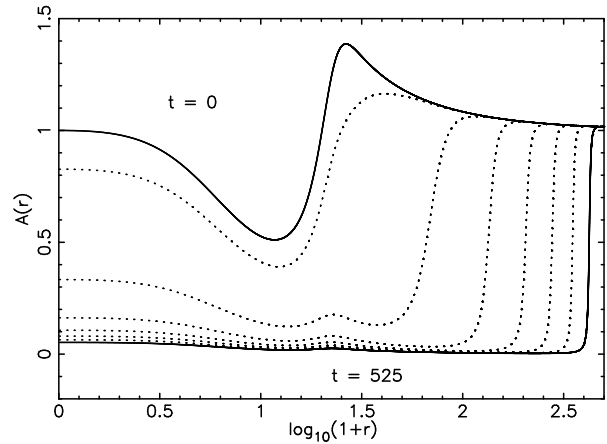,width=5.8cm,angle=270}
\caption{Evolution of the metric function $A$ for the perturbation
with \mbox{$\epsilon=+0.002$}.}
\label{fig:explosion_A} 
\end{figure}

\begin{figure}
\vspace{5mm}
\epsfig{file=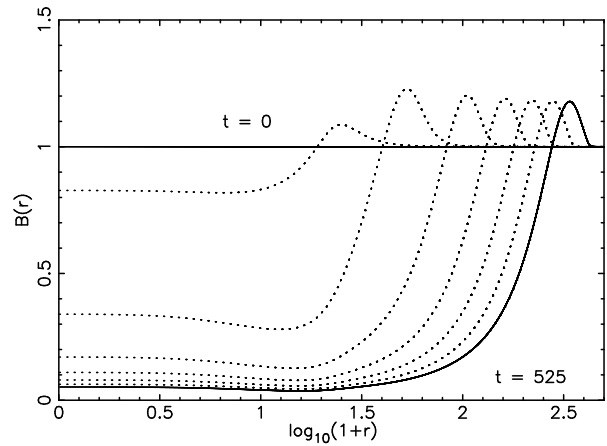,width=5.8cm,angle=270}
\caption{Same as Figure~\ref{fig:explosion_A} for
the metric function $B$.}
\label{fig:explosion_B} 
\end{figure}

Figures~\ref{fig:explosion_A} and~\ref{fig:explosion_B} show the
evolution of the metric functions $A$ and $B$ for this case (lines are
now separated in time by $\Delta t = 75$).  It is evident that the
dynamical evolution is completely different to the case described in
the previous section.  In the first place, there is no indication of
the slice stretching effect.  Moreover, in the evolution of the radial
metric $A$ it is clear that there is a wall moving outward. The wall
moves essentially at a uniform speed, even if this is not evident in
the log plot (this speed is approximately 1 in our units, which
coincides with the speed of light in the outer regions).  Inside this
wall the radial metric is collapsing to zero.  The angular metric is
also collapsing to zero in this region, but not as rapidly.

\begin{figure}[t]
\vspace{5mm}
\epsfig{file=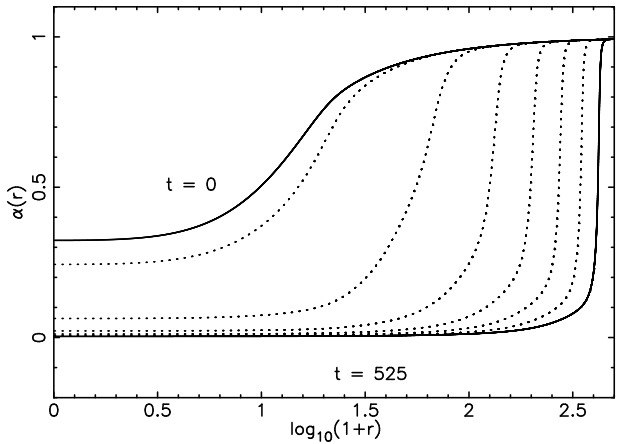,width=5.8cm,angle=270}
\caption{Same as Figure~\ref{fig:explosion_A} for
the lapse function $\alpha$.}
\label{fig:explosion_alpha} 
\end{figure}

The evolution of the lapse function is shown in
Fig.~\ref{fig:explosion_alpha}.  Again, the presence of an outward
moving wall is evident.  Inside the wall the lapse is collapsing to
zero, indicating that we are approaching a singularity.  However, in
this case no apparent horizon was located for the duration of the run.

\begin{figure}
\vspace{5mm}
\epsfig{file=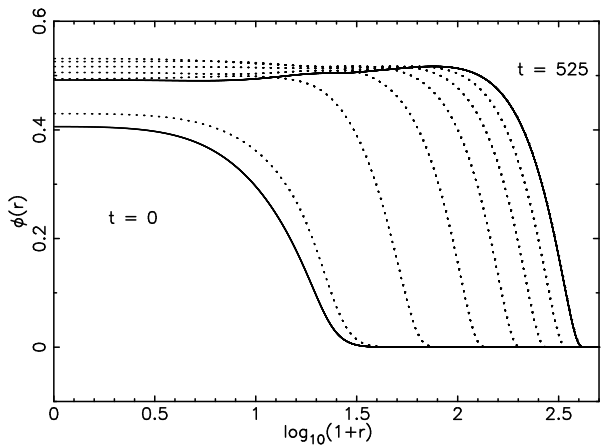,width=5.8cm,angle=270}
\caption{Evolution of the scalar field $\phi$ for the perturbation
with \mbox{$\epsilon=+0.002$}.  The scalar field is now moving toward
the true minimum of the potential at $\phi=0.5$ everywhere.}
\label{fig:explosion_phi} 
\end{figure}

The evolution of the scalar field can be seen in
Fig.~\ref{fig:explosion_phi}.  In contrast to the results of the
previous section, in this case the scalar field is moving toward the
true minimum of the potential at $\phi=0.5$.  Since this minimum
corresponds to a negative value of the potential, the interior of the
wall resembles an anti-de-Sitter spacetime, except for the fact that
the scalar field is not uniform.  Still, one would expect the
formation of a big-crunch type singularity in this region in a finite
proper time~\cite{Hertog03a,Alcubierre04b}.  However, because of the 
singularity avoiding properties of harmonic slicing, this singularity 
would only be reached after an infinite coordinate time.

\begin{figure}
\vspace{5mm}
\epsfig{file=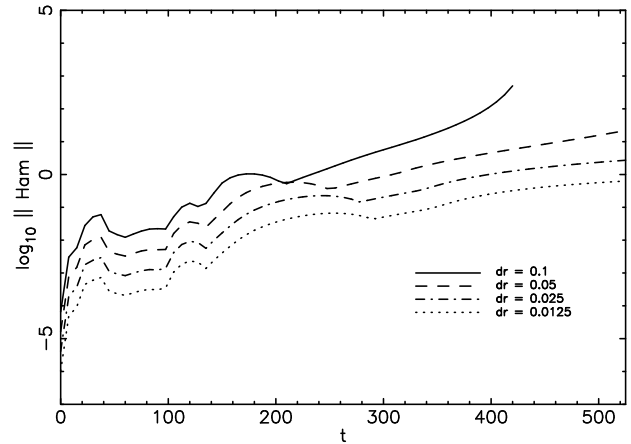,width=5.8cm,angle=270}
\caption{Logarithm of the root mean square of the hamiltonian
constraint as a function of time for the perturbation with
\mbox{$\epsilon=+0.002$}, for four different resolutions.}
\label{fig:explosion_conv} 
\end{figure}

Finally, Figure~\ref{fig:explosion_conv} shows the evolution of the
logarithm of the root mean square of the Hamiltonian constraint for
the same four resolutions.  Second order convergence is again evident.

The simulation seems to indicate that the perturbation has triggered a
global phase transition from the false vacuum in the exterior to the
true vacuum in the interior.  This phase transition propagates through
a domain wall that always moves outward, which implies that the
spacetime will never reach a stationary state.

%%%%%%%%%%%%%%%%%%%%%%
%%%   DISCUSSION   %%%
%%%%%%%%%%%%%%%%%%%%%%

\section{Discussion}
\label{sec:discussion}

We have considered the dynamical evolution of static self-gravitating
scalar solitons.  These soliton configurations arise for
self-interaction potentials $V(\phi)$ that have a local minimum for
which $V=0$, and a global minimum such that $V<0$.  This allows for
static, asymptotically flat solutions for which the scalar field
interpolates between the two minima.  The static configurations were
shown to be unstable~\cite{Nucamendi03}, and we have studied their 
response to two different types of perturbations differing in the 
sign of the perturbing term.

In one case the configuration is found to undergo gravitational
collapse, as the scalar field moves toward the local minimum of the
potential everywhere.  A small amount of scalar field is radiated
away, and the spacetime finally settles to a stationary Schwarzschild
BH.

For the other type of perturbation the scalar field ``explodes'',
triggering a global phase transition that propagates through an
outward moving domain wall.  This wall separates an inner region where
the scalar field moves toward the true minimum of the potential (the
true vacuum), and an outer region with the scalar field in the local
minimum.  As the true minimum has $V<0$, the inner region behaves in a
manner similar to anti-de-Sitter spacetime, and should produce a
big-crunch type singularity in a finite proper time.  The spacetime in
this case does not reach a stationary state, as the domain wall always
keeps moving outwards.  A question arises as to whether the
singularity that is forming in this case is naked, since there is no
evidence for an apparent horizon~\cite{Hertog03a,Alcubierre04b}.  
We believe that this is unlikely, as nothing special seems to happen 
during the evolution. More likely an event horizon does exist, but 
this is no BH in the standard sense since, first, no trapped surfaces 
form inside it, and second, a stationary state is never reached and 
the BH eventually swallows the whole spacetime.

%%%%%%%%%%%%%%%%%%%%%%%%%%%
%%%   ACKNOWLEDGMENTS   %%%
%%%%%%%%%%%%%%%%%%%%%%%%%%%

\acknowledgments

The authors wish to thank Bernd Reimann for the use of his code to
solve the maximal slicing condition in spherical symmetry.  We also
wish to acknowledge partial support from DGAPA-UNAM grants
No. IN112401 and No. IN122002, and Conacyt grant No. 32551-E.

%%%%%%%%%%%%%%%%%%%%
%%%   APPENDIX   %%%
%%%%%%%%%%%%%%%%%%%%

\appendix*
\section{}
\label{sec:appendix}

Recently, the need for having a hyperbolic system of evolution
equations has been stressed by several authors as a necessary
condition for the well posedness of the Cauchy initial data problem
(and also for the stability of the numerical evolution).  In order to
comply with this requirement we have re-written the ADM equations
together with the harmonic slicing condition as a first order system
of the form
\begin{equation}
\partial_t {\vec w} + \mathbb{M}^r \partial_r {\vec w} = {\vec S}\,\,\,,
\end{equation}
with ${\vec w}$ a first order variables vector
\begin{equation} 
{\vec w} := \Big( D_B, K_B, D_{\alpha}, K, U, \lambda \Big) \;,
\end{equation}
$\mathbb{M}^r$ the matrix
\begin{equation} 
\mathbb{M}^r = \left( \begin{array}{cccccc}
0          & 2 \alpha        & 0          & 0        & 0 & 0 \\ 
\alpha/2A  & 0               & 0          & 0        & 0 & 0 \\
0          & 0               & 0          & \alpha   & 0 & 0 \\
0          & 0               & \alpha / A & 0        & 0 & 0 \\
0          & 0               & 0          & 2 \alpha & 0 & 0 \\
0          & - 2 \alpha A /B & 0          & 0        & 0 & 0
\end{array} \right) \; ,
\end{equation}
and ${\vec S}$ source terms that include no derivatives (the variables
$\{\alpha,A,B\}$ evolve only through source terms and need not be
considered for the hyperbolicity analysis).

The eigenvalues of the matrix $\mathbb{M}^r$ turn out to be
\begin{eqnarray}
\lambda_{1,2} &=& \pm \frac{\alpha}{\sqrt{A}} \;, \\
\lambda_{3,4} &=& \pm \frac{\alpha}{\sqrt{A}} \;, \\
\lambda_{5,6} &=& 0 \;, 
\end{eqnarray}
with corresponding eigenvectors
\begin{eqnarray}
{\vec e}_{1,2} &=& \Big(\frac{B}{A}, \pm \frac{B}{2 A^{3/2}},0,0,0,-1\Big) 
\;, \\
{\vec e}_{3,4} &=& \Big(0,0,\frac{1}{2},\pm\frac{1}{2}\sqrt{\frac{1}{A}},1,0\Big)
\;, \\
{\vec e}_5 &=& \Big(0,0,0,0,1,0 \Big) \;, \\
{\vec e}_6 &=& \Big(0,0,0,0,0,1 \Big) \;,
\end{eqnarray}

Clearly, all eigenvalues are real.  Also, the eigenvector matrix has
determinant equal to $B^2/2 A^3 > 0$, showing the linear independence
of the eigenvectors.  The system of equations is therefore strongly
hyperbolic.

The above analysis applies to the case of harmonic slicing.  For
maximal slicing the argument is similar, except that in that case $K$
is no longer a dynamical variable ($K=0$), and the lapse is obtained
from an elliptic equation.  The reduced system of equations for
$\{D_B, K_B, U, \lambda\}$ is still strongly hyperbolic, and the full
system now has both an elliptic and a hyperbolic sector.

%%%%%%%%%%%%%%%%%%%%%%
%%%   REFERENCES   %%%
%%%%%%%%%%%%%%%%%%%%%%

\bibliographystyle{bibtex/apsrev}
\bibliography{bibtex/referencias}

\begin{thebibliography}{20}
\expandafter\ifx\csname natexlab\endcsname\relax\def\natexlab#1{#1}\fi
\expandafter\ifx\csname bibnamefont\endcsname\relax
  \def\bibnamefont#1{#1}\fi
\expandafter\ifx\csname bibfnamefont\endcsname\relax
  \def\bibfnamefont#1{#1}\fi
\expandafter\ifx\csname citenamefont\endcsname\relax
  \def\citenamefont#1{#1}\fi
\expandafter\ifx\csname url\endcsname\relax
  \def\url#1{\texttt{#1}}\fi
\expandafter\ifx\csname urlprefix\endcsname\relax\def\urlprefix{URL }\fi
\providecommand{\bibinfo}[2]{#2}
\providecommand{\eprint}[2][]{\url{#2}}

\bibitem[{\citenamefont{Nucamendi and Salgado}(2003)}]{Nucamendi03}
\bibinfo{author}{\bibfnamefont{U.}~\bibnamefont{Nucamendi}} \bibnamefont{and}
  \bibinfo{author}{\bibfnamefont{M.}~\bibnamefont{Salgado}},
  \bibinfo{journal}{Phys. Rev. D} \textbf{\bibinfo{volume}{68}},
  \bibinfo{pages}{044026} (\bibinfo{year}{2003}),
  \bibinfo{note}{gr-qc/0301062}.

\bibitem[{\citenamefont{Sudarsky}(1995)}]{Sudarsky95}
\bibinfo{author}{\bibfnamefont{D.}~\bibnamefont{Sudarsky}},
  \bibinfo{journal}{Class. Quantum Grav.} \textbf{\bibinfo{volume}{12}},
  \bibinfo{pages}{579} (\bibinfo{year}{1995}).

\bibitem[{\citenamefont{Bekenstein}(1995)}]{Bekenstein95}
\bibinfo{author}{\bibfnamefont{J.~D.} \bibnamefont{Bekenstein}},
  \bibinfo{journal}{Phys. Rev. D} \textbf{\bibinfo{volume}{51}},
  \bibinfo{pages}{R6608} (\bibinfo{year}{1995}).

\bibitem[{\citenamefont{Heusler}(1992)}]{Heusler92}
\bibinfo{author}{\bibfnamefont{M.}~\bibnamefont{Heusler}}, \bibinfo{journal}{J.
  Math. Phys.} \textbf{\bibinfo{volume}{33}}, \bibinfo{pages}{3497}
  (\bibinfo{year}{1992}).

\bibitem[{\citenamefont{Ashtekar et~al.}(2000)\citenamefont{Ashtekar, Beetle,
  Dreyer, Fairhurst, Krishnan, Lewandowski, and Wi\'sniewski}}]{Ashtekar00}
\bibinfo{author}{\bibfnamefont{A.}~\bibnamefont{Ashtekar}},
  \bibinfo{author}{\bibfnamefont{C.}~\bibnamefont{Beetle}},
  \bibinfo{author}{\bibfnamefont{O.}~\bibnamefont{Dreyer}},
  \bibinfo{author}{\bibfnamefont{S.}~\bibnamefont{Fairhurst}},
  \bibinfo{author}{\bibfnamefont{B.}~\bibnamefont{Krishnan}},
  \bibinfo{author}{\bibfnamefont{J.}~\bibnamefont{Lewandowski}},
  \bibnamefont{and}
  \bibinfo{author}{\bibfnamefont{J.}~\bibnamefont{Wi\'sniewski}},
  \bibinfo{journal}{Phys. Rev. Lett.} \textbf{\bibinfo{volume}{85}},
  \bibinfo{pages}{3564} (\bibinfo{year}{2000}).

\bibitem[{\citenamefont{Ashtekar et~al.}(2001)\citenamefont{Ashtekar, Corichi,
  and Sudarsky}}]{Ashtekar01}
\bibinfo{author}{\bibfnamefont{A.}~\bibnamefont{Ashtekar}},
  \bibinfo{author}{\bibfnamefont{A.}~\bibnamefont{Corichi}}, \bibnamefont{and}
  \bibinfo{author}{\bibfnamefont{D.}~\bibnamefont{Sudarsky}},
  \bibinfo{journal}{Class. Quantum. Grav.} \textbf{\bibinfo{volume}{18}},
  \bibinfo{pages}{919} (\bibinfo{year}{2001}).

\bibitem[{\citenamefont{hong Zhou and Straumann}(1991)}]{Zhou91}
\bibinfo{author}{\bibfnamefont{Z.}~\bibnamefont{hong Zhou}} \bibnamefont{and}
  \bibinfo{author}{\bibfnamefont{N.}~\bibnamefont{Straumann}},
  \bibinfo{journal}{Nucl. Phys. B} \textbf{\bibinfo{volume}{360}},
  \bibinfo{pages}{180} (\bibinfo{year}{1991}).

\bibitem[{\citenamefont{Arnowitt et~al.}(1962)\citenamefont{Arnowitt, Deser,
  and Misner}}]{Arnowitt62}
\bibinfo{author}{\bibfnamefont{R.}~\bibnamefont{Arnowitt}},
  \bibinfo{author}{\bibfnamefont{S.}~\bibnamefont{Deser}}, \bibnamefont{and}
  \bibinfo{author}{\bibfnamefont{C.~W.} \bibnamefont{Misner}}, in
  \emph{\bibinfo{booktitle}{Gravitation: An Introduction to Current Research}},
  edited by \bibinfo{editor}{\bibfnamefont{L.}~\bibnamefont{Witten}}
  (\bibinfo{publisher}{John Wiley}, \bibinfo{address}{New York},
  \bibinfo{year}{1962}), pp. \bibinfo{pages}{227--265}.

\bibitem[{\citenamefont{York}(1979)}]{York79}
\bibinfo{author}{\bibfnamefont{J.}~\bibnamefont{York}}, in
  \emph{\bibinfo{booktitle}{Sources of Gravitational Radiation}}, edited by
  \bibinfo{editor}{\bibfnamefont{L.}~\bibnamefont{Smarr}}
  (\bibinfo{publisher}{Cambridge University Press},
  \bibinfo{address}{Cambridge, England}, \bibinfo{year}{1979}).

\bibitem[{\citenamefont{Alcubierre and Gonz\'alez}(2004)}]{Alcubierre04a}
\bibinfo{author}{\bibfnamefont{M.}~\bibnamefont{Alcubierre}} \bibnamefont{and}
  \bibinfo{author}{\bibfnamefont{J.~A.} \bibnamefont{Gonz\'alez}}
  (\bibinfo{year}{2004}), \bibinfo{note}{gr-qc/0401113}.

\bibitem[{\citenamefont{Bona et~al.}(1995)\citenamefont{Bona, Mass{\'o},
  Seidel, and Stela}}]{Bona94b}
\bibinfo{author}{\bibfnamefont{C.}~\bibnamefont{Bona}},
  \bibinfo{author}{\bibfnamefont{J.}~\bibnamefont{Mass{\'o}}},
  \bibinfo{author}{\bibfnamefont{E.}~\bibnamefont{Seidel}}, \bibnamefont{and}
  \bibinfo{author}{\bibfnamefont{J.}~\bibnamefont{Stela}},
  \bibinfo{journal}{Phys. Rev. Lett.} \textbf{\bibinfo{volume}{75}},
  \bibinfo{pages}{600} (\bibinfo{year}{1995}), \bibinfo{note}{gr-qc/9412071}.

\bibitem[{\citenamefont{Bona et~al.}(1997)\citenamefont{Bona, Mass{\'o},
  Seidel, and Stela}}]{Bona97a}
\bibinfo{author}{\bibfnamefont{C.}~\bibnamefont{Bona}},
  \bibinfo{author}{\bibfnamefont{J.}~\bibnamefont{Mass{\'o}}},
  \bibinfo{author}{\bibfnamefont{E.}~\bibnamefont{Seidel}}, \bibnamefont{and}
  \bibinfo{author}{\bibfnamefont{J.}~\bibnamefont{Stela}},
  \bibinfo{journal}{Phys. Rev. D} \textbf{\bibinfo{volume}{56}},
  \bibinfo{pages}{3405} (\bibinfo{year}{1997}), \bibinfo{note}{gr-qc/9709016}.

\bibitem[{\citenamefont{Alcubierre}(2003)}]{Alcubierre02b}
\bibinfo{author}{\bibfnamefont{M.}~\bibnamefont{Alcubierre}},
  \bibinfo{journal}{Class. Quantum Grav.} \textbf{\bibinfo{volume}{20}},
  \bibinfo{pages}{607} (\bibinfo{year}{2003}), \bibinfo{note}{gr-qc/0210050}.

\bibitem[{\citenamefont{Smarr and York}(1978)}]{Smarr78b}
\bibinfo{author}{\bibfnamefont{L.}~\bibnamefont{Smarr}} \bibnamefont{and}
  \bibinfo{author}{\bibfnamefont{J.}~\bibnamefont{York}},
  \bibinfo{journal}{Phys. Rev. D} \textbf{\bibinfo{volume}{17}},
  \bibinfo{pages}{2529} (\bibinfo{year}{1978}).

\bibitem[{\citenamefont{Hertog et~al.}(2003)\citenamefont{Hertog, Horowitz, and
  Maeda}}]{Hertog03a}
\bibinfo{author}{\bibfnamefont{T.}~\bibnamefont{Hertog}},
  \bibinfo{author}{\bibfnamefont{G.~T.} \bibnamefont{Horowitz}},
  \bibnamefont{and} \bibinfo{author}{\bibfnamefont{K.}~\bibnamefont{Maeda}}
  (\bibinfo{year}{2003}), \bibinfo{note}{gr-qc/0307102}.

\bibitem[{\citenamefont{Alcubierre et~al.}(2004)\citenamefont{Alcubierre,
  Gonz\'alez, Salgado, and Sudarsky}}]{Alcubierre04b}
\bibinfo{author}{\bibfnamefont{M.}~\bibnamefont{Alcubierre}},
  \bibinfo{author}{\bibfnamefont{J.~A.} \bibnamefont{Gonz\'alez}},
  \bibinfo{author}{\bibfnamefont{M.}~\bibnamefont{Salgado}}, \bibnamefont{and}
  \bibinfo{author}{\bibfnamefont{D.}~\bibnamefont{Sudarsky}}
  (\bibinfo{year}{2004}), \bibinfo{note}{gr-qc/0402045}.

\bibitem[{\citenamefont{Alcubierre et~al.}(2000)\citenamefont{Alcubierre,
  Br\"ugmann, Dramlitsch, Font, Papadopoulos, Seidel, Stergioulas, and
  Takahashi}}]{Alcubierre99d}
\bibinfo{author}{\bibfnamefont{M.}~\bibnamefont{Alcubierre}},
  \bibinfo{author}{\bibfnamefont{B.}~\bibnamefont{Br\"ugmann}},
  \bibinfo{author}{\bibfnamefont{T.}~\bibnamefont{Dramlitsch}},
  \bibinfo{author}{\bibfnamefont{J.}~\bibnamefont{Font}},
  \bibinfo{author}{\bibfnamefont{P.}~\bibnamefont{Papadopoulos}},
  \bibinfo{author}{\bibfnamefont{E.}~\bibnamefont{Seidel}},
  \bibinfo{author}{\bibfnamefont{N.}~\bibnamefont{Stergioulas}},
  \bibnamefont{and}
  \bibinfo{author}{\bibfnamefont{R.}~\bibnamefont{Takahashi}},
  \bibinfo{journal}{Phys. Rev. D} \textbf{\bibinfo{volume}{62}},
  \bibinfo{pages}{044034} (\bibinfo{year}{2000}),
  \bibinfo{note}{gr-qc/0003071}.

\bibitem[{\citenamefont{Hawking and Ellis}(1973)}]{Hawking73}
\bibinfo{author}{\bibfnamefont{S.~W.} \bibnamefont{Hawking}} \bibnamefont{and}
  \bibinfo{author}{\bibfnamefont{G.~F.~R.} \bibnamefont{Ellis}},
  \emph{\bibinfo{title}{The Large Scale Structure of Spacetime}}
  (\bibinfo{publisher}{Cambridge University Press},
  \bibinfo{address}{Cambridge, England}, \bibinfo{year}{1973}).

\bibitem[{\citenamefont{Wald}(1984)}]{Wald84}
\bibinfo{author}{\bibfnamefont{R.~M.} \bibnamefont{Wald}},
  \emph{\bibinfo{title}{General Relativity}} (\bibinfo{publisher}{The
  University of Chicago Press}, \bibinfo{address}{Chicago, U.S.A.},
  \bibinfo{year}{1984}).

\bibitem[{\citenamefont{York}(1989)}]{York89}
\bibinfo{author}{\bibfnamefont{J.}~\bibnamefont{York}}, in
  \emph{\bibinfo{booktitle}{Frontiers in Numerical Relativity}}, edited by
  \bibinfo{editor}{\bibfnamefont{C.}~\bibnamefont{Evans}},
  \bibinfo{editor}{\bibfnamefont{L.}~\bibnamefont{Finn}}, \bibnamefont{and}
  \bibinfo{editor}{\bibfnamefont{D.}~\bibnamefont{Hobill}}
  (\bibinfo{publisher}{Cambridge University Press},
  \bibinfo{address}{Cambridge, England}, \bibinfo{year}{1989}), pp.
  \bibinfo{pages}{89--109}.

\end{thebibliography}

%%%%%%%%%%%%%%%
%%%   END   %%%
%%%%%%%%%%%%%%%

\end{document}